\documentclass[conference]{IEEEtran}
\ifCLASSINFOpdf
\else
\fi
\ifCLASSOPTIONcompsoc
  \usepackage[caption=false,font=normalsize,labelfont=sf,textfont=sf]{subfig}
\else
  \usepackage[caption=false,font=footnotesize]{subfig}
\fi
\usepackage[cmex10]{amsmath}
\ifCLASSINFOpdf
   \usepackage[pdftex]{graphicx}
\else
   \usepackage[dvips]{graphicx}
\fi

\usepackage{amssymb}
\usepackage{epstopdf}

\usepackage{graphicx}
\usepackage[caption=false]{subfig}
\usepackage{enumerate}

\usepackage{verbatim}

\usepackage{microtype}
\DisableLigatures{encoding = *, family = * }

\usepackage[nolist]{acronym}
\begin{acronym}
	\acro{AAA}{Authentication, Authorization and Accounting}
	\acro{ACL}{Access Control List}
	\acro{AKI}{Accountable Key Infrastructure}
	\acro{API}{Application Programming Interface}
	\acro{ASS}{Anonymity Set Size}
    \acro{BSM}{Basic Safety Message}
    \acro{BYOD}{Bring Your Own Device}
	\acro{C2C-CC}{Car2Car Communication Consortium}
	\acro{C2C}{Car-to-Car}
	\acro{C2I}{Car-to-Infrastructure}
	\acro{CA}{Certification Authority}
	\acro{CN}{Common Name}
	\acro{CAM}{Cooperative Awareness Message}
	\acro{CAMP VSC3}{Crash Avoidance Metrics Partnership Vehicle Safety Consortium}
	\acro{CIA}{Confidentiality, Integrity and Availability}
	\acro{CRL}{Certificate Revocation List}
	\acro{CDN}{Content Delivery Network}
	\acro{COCA}{Cornell OnLine Certification Authority}
	\acro{CSR}{Certificate Signing Request}
	\acro{DAA}{Direct Anonymous Attestation}
	\acro{DDoS}{Distributed Denial of Service}
	\acro{DDH}{Decisional Diffie-Helman}
	\acro{DENM}{Decentralized Environmental Notification Message}
	\acro{DHT}{Distributed Hash Table}
	\acro{DL/ECIES}{Discrete Logarithm and Elliptic Curve Integrated Encryption Scheme}
	\acro{DoS}{Denial of Service}
	\acro{DoT}{Department of Transportation}
	\acro{DPA}{Data Protection Agency}
	\acro{DSRC}{Dedicated Short Range Communication}
	\acro{DSS}{Digital Signature Standard}
	\acro{ECU}{Electronic Control Unit}
	\acro{EDR}{Event Data Recorder}
	\acro{ETSI}{European Telecommunications Standards Institute}
	\acro{ECDSA}{Elliptic Curve Digital Signature Algorithm}
	\acro{ECC}{Elliptic Curve Cryptography}
	\acro{EVITA}{E-safety Vehicle Intrusion protected Applications}
	\acro{FOT}{Field Operational Testing}
	\acro{FPGA}{Field-Programmable Gate Array}
	\acro{GPA}{Global Passive Adversary}
	\acro{GN}{GeoNetworking}
	\acro{GS-VLR}{Group Signatures with Verifier Local Revocation}
	\acro{GS}{Group Signatures}
	\acro{GM}{Group Manager}
	\acro{GBA}{Generic Bootstrapping Architecture}
	\acro{GUI}{Graphic User Interface}
	\acro{HSM}{Hardware Security Module}
	\acro{HTTP}{Hypertext Transfer Protocol}
	\acro{IEEE}{Institute of Electrical and Electronics Engineers}
	\acro{IETF}{Internet Engineering Task Force}
	\acro{IoT}{Internet of Things}
	\acro{ITS}{Intelligent Transport System}
	\acro{IT}{Information Technologies}
	\acro{IMSI}{International Mobile Subscriber Identity}
	\acro{IMEI}{International Mobile Station Equipment Identity}
	\acro{IdP}{Identity Provider}
	\acro{IDS}{Intrusion Detection System}
	\acro{ISP}{Internet Service Provider}
	\acro{LEA}{Law Enforcement Agency}
	\acro{LCPP}{Lightweight Conditional Privacy Preservation}
	\acro{LTC}{Long Term Certificate}
	\acro{LTCA}{Long Term \acs{CA}}
	\acro{H-LTCA}{Home-LTCA}
	\acro{F-LTCA}{Foreign-LTCA}
	\acro{LDAP}{Lightweight Directory Access Protocol}
	\acro{LBS}{Location Based Service}
	\acro{LTE}{Long Term Evolution}
	\acro{LuST}{Luxembourg SUMO Traffic}
	\acro{MAC}{Message Authentication Code}
	\acro{MCA}{Message \ac{CA}}
	\acro{MEA}{Misbehavior Evaluation Authority}
	\acro{OBU}{On-Board Unit}
	\acro{OEM}{Original Equipment Manufacturer}
	\acro{OCSP}{Online Certificate Status Protocol}
	\acro{PCA}{Pseudonym \acs{CA}}
	\acro{PDP}{Policy Decision Point}
	\acro{PEP}{Policy Enforcement Point}
	\acro{PIR}{Private Information Retrieval}
	\acro{PKC}{Public Key Cryptography}
	\acro{PKCS}{Public Key Cryptosystem}
	\acro{PKI}{Public-Key Infrastructure}
	\acro{PRECIOSA}{Privacy Enabled Capability in Co-operative Systems and Safety Applications}
	\acro{PRESERVE}{Preparing Secure Vehicle-to-X Communication Systems}
	\acro{P2P}{peer-to-peer}
	\acro{PS}{Participatory Sensing}
	\acro{RA}{Resolution Authority}
	\acro{REST}{Representational State Transfer}
	\acro{RBAC}{Role Based Access Control}
	\acro{RCA}{Root \acs{CA}}
	\acro{RSU}{Roadside Unit}
	\acro{RHyTHM}{RHyTHM}
	\acro{SAML}{Security Assertion Markup Language}
	\acro{SAS}{Sample Aggregation Service}
	\acro{SCMS}{Security Credential Management System}
	\acro{SCORE@F}{Système COopératif Routier Expérimental Français}
	\acro{SDSI}{Simple Distributed Security Infrastructure}
	\acro{SRAAC}{Secure Revocable Anonymous Authenticated Inter-Vehicle Communication}
	\acro{SeVeCom}{Secure Vehicle Communication}
	\acro{SIT}{Sichere Informationstechnologie}
	\acro{SLC}{Short-Lived Certificate}
	\acro{SoA}{Service-oriented-Approach}
	\acro{SIFS}{Short Inter Frame Space}
	\acro{SSO}{Single-Sign-On}
	\acro{SSL}{Secure Sockets Layer}
	\acro{SOAP}{Simple Object Access Protocol}
	\acro{TACK}{Temporary Anonymous Certified Key}
	\acro{TS}{Task Service}
	\acro{TLS}{Transport Layer Security}
	\acro{TPM}{Trusted Platform Module}
	\acro{TTP}{Trusted Third Party}
	\acro{TVR}{Ticket Validation Repository}
	\acro{URI}{Uniform Resource Identifier}
	\acro{UML}{Unified Modeling Language}
	\acro{VANET}{Vehicular Ad-hoc Network}
	\acro{V2I}{Vehicle-to-Infrastructure}
	\acro{V2V}{Vehicle-to-Vehicle}
	\acro{V2X}{\ac{V2V} and \ac{V2I}}
	\acro{VC}{Vehicular Communication}
	\acro{VM}{Virtual Machine}
	\acro{VSS}{\ac{VC} Security Subsystem}
	\acro{WAVE}{Wireless Access in Vehicular Environments}
	\acro{WSDL}{Web Services Discovery Language}
	\acro{W3C}{World Wide Web Consortium}
	\acro{V}{Vehicle}
	\acro{VANET}{Vehicular Ad-hoc Network}
	\acro{VLR}{Verifier-Local Revocation}
	\acro{VPKI}{Vehicular Public-Key Infrastructure}
	\acro{VM}{Virtual Machine}
	\acro{WS}{Web Service}
	\acro{WoT}{Web of Trust}
	\acro{WSACA}{\ac{WAVE} Service Advertisement \ac{CA}}
	\acro{XML}{Extensible Markup Language}
	\acro{XACML}{eXtensible Access Control Markup Language}
	\acro{ZKP}{Zero Knowledge Proof}
	\acro{3G}{3rd Generation}
\end{acronym}

\usepackage{makecell}

\usepackage{xcolor,colortbl}

\definecolor{Gray}{gray}{0.85}
\definecolor{LightCyan}{rgb}{0.88,1,1}
\newcolumntype{a}{>{\columncolor{Gray}}c}
\newcolumntype{b}{>{\columncolor{white}}c}

\usepackage{amssymb}
\usepackage{epstopdf}

\usepackage{graphicx}
\usepackage[caption=false]{subfig}

\usepackage{enumerate}
\usepackage{slashbox}
\usepackage{amsmath}
\usepackage{verbatim}

\usepackage{microtype}
\DisableLigatures{encoding = *, family = * }

\usepackage{makecell}

\usepackage{arydshln}

\usepackage{xcolor}

\usepackage{algpseudocode}

\usepackage{setspace}

\errorcontextlines\maxdimen

\makeatletter
\newcommand*{\algrule}[1][\algorithmicindent]{\makebox[#1][l]{\hspace*{.5em}\thealgruleextra\vrule height \thealgruleheight depth \thealgruledepth}}%
\newcommand*{\thealgruleextra}{}
\newcommand*{\thealgruleheight}{.75\baselineskip}
\newcommand*{\thealgruledepth}{.25\baselineskip}

\newcount\ALG@printindent@tempcnta
\def\ALG@printindent{%
	\ifnum \theALG@nested>0
	\ifx\ALG@text\ALG@x@notext
	\else
	\unskip
	\addvspace{-1pt}
	\ALG@printindent@tempcnta=1
	\loop
	\algrule[\csname ALG@ind@\the\ALG@printindent@tempcnta\endcsname]%
	\advance \ALG@printindent@tempcnta 1
	\ifnum \ALG@printindent@tempcnta<\numexpr\theALG@nested+1\relax
	\repeat
	\fi
	\fi
}%
\usepackage{etoolbox}
\patchcmd{\ALG@doentity}{\noindent\hskip\ALG@tlm}{\ALG@printindent}{}{\errmessage{failed to patch}}
\makeatother

\newbox\statebox
\newcommand{\myState}[1]{%
	\setbox\statebox=\vbox{#1}%
	\edef\thealgruleheight{\dimexpr \the\ht\statebox+1pt\relax}%
	\edef\thealgruledepth{\dimexpr \the\dp\statebox+1pt\relax}%
	\ifdim\thealgruleheight<.75\baselineskip
	\def\thealgruleheight{\dimexpr .75\baselineskip+1pt\relax}%
	\fi
	\ifdim\thealgruledepth<.25\baselineskip
	\def\thealgruledepth{\dimexpr .25\baselineskip+1pt\relax}%
	\fi
	\State #1%
	\def\thealgruleheight{\dimexpr .75\baselineskip+1pt\relax}%
	\def\thealgruledepth{\dimexpr .25\baselineskip+1pt\relax}%
}

\usepackage{footnote}

\usepackage{rotating}

\usepackage{soul}

\usepackage{algorithm}

\usepackage{tikz}
\usetikzlibrary{shadows,positioning,calc}
\tikzset{multiple/.style = {double copy shadow={shadow xshift=1ex,shadow
			yshift=-1.5ex,draw=black!30},fill=white,draw=black,thick,minimum height = 1cm,minimum
		width=2cm},
	ordinary/.style = {rectangle,draw,thick,minimum height = 1cm,minimum width=2cm}}

\usetikzlibrary{decorations.pathreplacing}

\usepackage{algpseudocode}

\makeatletter
\renewcommand{\ALG@beginalgorithmic}{\scriptsize}
\makeatother

\usepackage{afterpage}
\newlength{\oldtextfloatsep}

\begin{document}
%
\title{\acl{RHyTHM}: A Randomized Hybrid Scheme \\ To Hide in the Mobile Crowd}


\author{\IEEEauthorblockN{Mohammad Khodaei, Andreas Messing, and Panos Papadimitratos} 
\IEEEauthorblockA{Networked Systems Security Group, KTH Royal Institute of Technology, Stockholm, Sweden \\
\emph{\{khodaei, amessing, papadim\}}@kth.se}}

\maketitle

\begin{abstract}

Any on-demand pseudonym acquisition strategy is problematic should the connectivity to the credential management infrastructure be intermittent. If a vehicle runs out of pseudonyms with no connectivity to refill its pseudonym pool, one solution is the \emph{on-the-fly} generation of pseudonyms, e.g., leveraging anonymous authentication. However, such a vehicle would stand out in the crowd: one can simply distinguish pseudonyms, thus signed messages, based on the pseudonym issuer signature, link them and track the vehicle. To address this challenge, we propose a randomized hybrid scheme, \acl{RHyTHM}, to enable vehicles to remain operational when disconnected without compromising privacy: vehicles with valid pseudonyms help others to enhance their privacy by randomly joining them in using \emph{on-the-fly} \emph{self-certified} pseudonyms along with aligned lifetimes. This way, the privacy of disconnected users is enhanced with a reasonable computational overhead. 

\end{abstract}




%
\IEEEpeerreviewmaketitle

\section{Introduction}
\label{sec:introduction}

\acresetall

In \ac{VC} systems, vehicles beacon \acp{CAM} and \acp{DENM} periodically at a high rate in order to provide cooperative awareness. \ac{V2X} communication is protected with the help of public key cryptography: a set of short-term anonymous credentials, i.e., \emph{pseudonyms}, are provided to each vehicle by the \ac{VPKI}, e.g.,~\cite{khodaei2016Secmace}. Thus, vehicles switch from one pseudonym to another for message unlinkability as pseudonyms are inherently unlinkable. 

One can provide vehicles with valid pseudonyms for a long period, e.g., 25 years~\cite{kumar2017binary}. However, extensive preloading with millions of pseudonyms per vehicle for such a long period is computationally costly, inefficient in utilization and cumbersome in revocation~\cite{khodaei2015VTMagazine}. On the contrary, several proposals suggest more frequent Vehicle-to-\ac{VPKI} interactions, namely \emph{on-demand} schemes, e.g.,~\cite{khodaei2016Secmace, schaub2010v}. This strategy is more efficient in terms of pseudonym utilization and revocation and more effective in fending off misbehavior. However, the more frequent the interaction with the \ac{VPKI}, the more dependent vehicles are on connectivity. This may hurt Vehicle-to-\ac{VPKI} connectivity on intermittent coverage of sparsely-deployed \acp{RSU}, or highly overloaded existing cellular infrastructure. Thus, it is necessary to ensure that any vehicle at any time can continue its operation securely without harming privacy, even if the \ac{VPKI} is not reachable or available for other reasons, e.g., during a \ac{DoS} attack~\cite{khodaei2016Secmace}.

Obviously, signing \acp{CAM} with the private keys corresponding to expired pseudonyms or the long-term certificate, is insecure and harm user privacy, i.e., messages are trivially linkable. \emph{On-the-fly} generation of pseudonyms, the \emph{hybrid} scheme~\cite{calandriello2011performance}, using other anonymous authentication primitives, i.e., group signatures~\cite{boneh2004short}, is a promising alternative. Each vehicle is equipped with a group public key, common among all the group members, along with a distinct group signing key. In order to generate on-the-fly pseudonyms, each vehicle generates a pair of public/private keys and signs the public key using the group signing key instead of having a pseudonym signed by the corresponding \ac{CA}. This essentially eliminates the need to request pseudonyms from the \ac{VPKI} entities, especially when the latter is unreachable. This provides authenticity, integrity, accountability, and non-repudiation. Furthermore, a node can be evicted from the system if it deviated from the system security policy. 

If only a few vehicles use their self-certified pseudonyms while the rest of the vehicles rely on the \ac{VPKI}-provided pseudonyms, the \emph{baseline} scheme, they would \emph{``stand out in a crowd''}: one can simply distinguish the pseudonyms, thus the pseudonymously signed messages, based on the pseudonym issuer's signature. Moreover, the self-certified pseudonyms lifetimes are not aligned with each other and the global system time, i.e., the \ac{VPKI} clock. As a result, all the vehicles in a region will be transmitting under pseudonyms which are distinguishable based on their timing information~\cite{khodaei2016Secmace}.

To address this challenge, we propose a cooperative and adaptive scheme, \acl{RHyTHM}, to mitigate this privacy issue: a vehicle with no valid \ac{VPKI}-provided pseudonyms initiates \acl{RHyTHM} protocol by setting a flag in the upcoming \acp{CAM}. Neighboring vehicles with \ac{VPKI}-provided pseudonyms randomly opt in to utilize their self-certified pseudonyms with the probability of $r$ in upcoming pseudonym updates. \acl{RHyTHM} enhances the privacy of users running out of pseudonyms at the cost of reasonable processing overhead for neighboring vehicles. This ensures the operation of every legitimate vehicle without harming user privacy even if the infrastructure fails to provide them credentials. 

In the rest of the paper, we describe our system and adversarial model (Sec.~\ref{sec:adversarial-model}), present our scheme (Sec.~\ref{sec:rhythm}) and security and privacy analysis (Sec.~\ref{sec:security-analysis}). We provide the performance evaluation of our scheme (Sec.~\ref{sec:evaluation}) before conclusion (Sec.~\ref{sec:conclusions}). 

\vspace{-1em}
\section{System and Adversarial Model}
\label{sec:adversarial-model}

We assume a \ac{VPKI} with distinct entities and roles~\cite{khodaei2016Secmace}: the \ac{LTCA} is responsible for vehicles registration in a domain~\cite{khodaei2015VTMagazine}; the \ac{PCA} issues pseudonyms for the registered vehicles; and the \ac{RA} is able to initiate a process to resolve a pseudonym of a misbehaving vehicle. Furthermore, a \ac{GM} enables any legitimate vehicle to sign a message on behalf of the group without disclosing its actual identity. Upon registration of a vehicle by the \ac{LTCA} in the bootstrapping phase, each vehicle is provided with an \emph{anonymous ticket}, with which the \ac{GM} registers the vehicle, thus authorizes it to anonymously operate in some circumstances, e.g., the \ac{VPKI} is unreachable.

We consider external and internal adversaries that try to harm or abuse \acl{RHyTHM}. External adversaries could sign messages with fake private keys. Internal adversaries could initiate \acl{RHyTHM} protocol continuously for two purposes: (i) to be provided with multiple simultaneously valid pseudonyms, thus performing Sybil-based~\cite{douceur2002sybil} attacks; (ii) to compromise the availability of neighboring vehicles by incurring extra workload towards \ac{DoS} attacks. Moreover, a global observer, e.g., an \emph{honest-but-curious} \ac{VPKI} entity~\cite{khodaei2016Secmace}, might be tempted to link the \ac{VPKI}-provided pseudonyms to the self-certified ones to infer user sensitive information towards harming user privacy. 


\section{\acl{RHyTHM} Operation}
\label{sec:rhythm}

In order to achieve full unlinkability, we assume that a universally fixed interval, $\Gamma$, is specified in a region~\cite{khodaei2015VTMagazine} and all pseudonyms in that region are issued with the lifetime aligned with the global system time, i.e., the \ac{VPKI} clock. As a result of this policy, at any point in time, all the vehicles transmit using pseudonyms indistinguishable, from one another, thanks to this time alignment. This essentially eliminates any distinction among pseudonym sets of different vehicles, thus achieving user privacy protection. We refer readers to~\cite{khodaei2016Secmace} for further details. The \ac{OBU} \emph{``decides''} when to trigger the pseudonym acquisition process based on various parameters~\cite{khodaei2016evaluating}. This can happen even within the lifetime of the last single valid pseudonym should the connectivity to the \ac{VPKI} entities be reliable. However, if the \ac{VPKI} entities are out of reach for any reason, the \ac{OBU} initiates the \acl{RHyTHM} protocol to use its self-certified pseudonyms during the next pseudonym update. If the \ac{OBU} has no valid \ac{VPKI}-provided pseudonyms, it initiates \acl{RHyTHM} protocol with its self-certified pseudonym. Table~\ref{table:crl-dis-protocols-notation} summarizes notation used in the protocol.

The vehicle, $V$, generates multiple \ac{ECDSA} key pairs and aligns the validity intervals with the known \ac{VPKI} clock (steps 1\textendash7). The \ac{OBU} does not need to be fully synchronized with the \ac{VPKI} clock; it simply aligns the pseudonyms lifetimes, $\tau_{P}$, in the continuation of its last valid \ac{VPKI}-provided pseudonym. In case of having \ac{VPKI}-provided pseudonyms from a distant past and being unable to be synchronized by any other means, the \ac{OBU} aligns the self-certified pseudonyms based on the pseudonym information, piggybacked in neighbors \acp{CAM}. This eliminates any distinction among self-certified pseudonyms (signed by the group signing key, $gsk_{v}$); moreover, the anonymity set becomes equal to the number of vehicles with self-certified pseudonyms. Finally, $V$ signs them using the $gsk_{v}$. It then piggybacks \acp{CAM} to explicitly inform its neighbors of initializing the \acl{RHyTHM} protocol for the next pseudonym update (steps 8\textendash10). Upon reception of a \acl{RHyTHM} initiation query, the neighboring vehicles check if the \ac{VPKI} entities are indeed out of reach. Having had the same viewpoint on the \ac{VPKI} reachability, they explicitly set the \acl{RHyTHM} flag in the upcoming \acp{CAM} to inform their neighbors, thus epidemically distributing the message. This ensures the distribution of the \acl{RHyTHM} initiation query.

\begin{table} [t!] 
	\vspace{-1.75em}
	\caption{Notation used in the protocols}
	\vspace{-0.5em}
	\centering
	\resizebox{0.475\textwidth}{!}
	{
		\renewcommand{\arraystretch}{1.1}
		\begin{tabular}{l | *{1}{c} r}
			\hline 
			$(K^i_v, k^i_v)$ & \shortstack{pseudonymous public/private keys, corresponding to current pseudonym} \\\hline 
			$gsk_{v}$ & group signing key \\ \hline 
			$t_{now}, t_s, t_e$ & fresh/current, starting, and ending timestamps \\\hline
			$Sign(key, msg)$ & signing a message with private key or group signing key \\\hline
			$(msg)_{\sigma_{v}}, (msg)_{\Sigma_{gsk_{v}}}$ & a signed message with $k^i_v$ or $gsk_{v}$ \\
			\hline
		\end{tabular}
		\renewcommand{\arraystretch}{1}
		\label{table:crl-dis-protocols-notation}
	}
	\vspace{-1.71em}
\end{table}

\setlength{\textfloatsep}{0pt}
\begin{algorithm} [t!]
	\floatname{algorithm}{}
	\caption{{\footnotesize \acl{RHyTHM} Initiation Protocol}} 
	\label{protocol:rhythm-initiation-algorithm}
	\algloop{For}{}
	\algblock{Begin}{End}
	\begin{algorithmic}[1]
		\Procedure{RHyTHMInit}{$t_{s}, t_{e}$} 
		\For{i:=1 to \textbf{n}}{}
		\Begin
		\State $\text{Generate}(K^{i}_{v}, k^i_v)$
		\State $\zeta \gets (K^{i}_{v}, t^{i}_{s}, t^{i}_{e})$
		\State $(K^i_v)_{\Sigma_{k^i_v}} \gets \text{Sign}(gsk_v, \zeta)$
		\End
		\State $Flag_{rhythm} \gets True$
		\State $CAM \gets \{Fields, Flag_{rhythm}, t_{now}\}$
		\State $(CAM)_{\sigma_{k^i_v}} \gets \text{Sign}(CAM, K^{i}_{v})$
		\EndProcedure
	\end{algorithmic}
\end{algorithm}
\afterpage{\global\setlength{\textfloatsep}{\oldtextfloatsep}}

\begin{figure} [!t] 
	\vspace{-2.05em}
	\begin{center}
		\centering
		\includegraphics[trim=0cm 0cm 0cm 0cm, clip=true, totalheight=0.1001\textheight,angle=0,keepaspectratio]{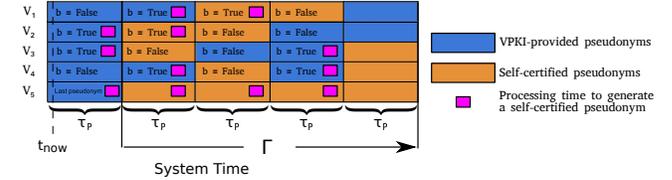}
		\vspace{-1em}
		\caption{\acl{RHyTHM} overview: if $b=True$, the vehicle will utilize its self-certified pseudonym; otherwise, it relies on its \ac{VPKI}-provided pseudonym.}
		\label{fig:rhythm-scheme-overview-with-pseudonym-updates}
	\end{center}
	\vspace{-0.5em}
\end{figure}

Fig.~\ref{fig:rhythm-scheme-overview-with-pseudonym-updates} illustrates five vehicles, out of which $V_{5}$ runs out of pseudonyms. It initiates the \acl{RHyTHM} protocol by setting the \acl{RHyTHM} flag in the upcoming \acp{CAM}. Neighboring vehicles, i.e., $V_{1}$\textendash$V_{4}$, randomly opt in to utilize their self-certified pseudonyms with probability $r$ in the first pseudonym update. $V_{2}$ and $V_{3}$ \emph{``decide''} to switch to utilize their self-certified pseudonyms, thus, they generate a pair of keys, align the validity interval with the global system time, and sign them with $gsk$. For the second pseudonym update in $\Gamma$, only $V_{3}$ \emph{``opted in''} to use its \ac{VPKI}-provided pseudonym while the rest of vehicles \emph{``decided''} to utilize their self-certified pseudonyms. $V_{5}$ is the only vehicle that uses its self-certified pseudonyms during the entire $\Gamma$ period while other vehicles randomly opt in to use either of the two. As vehicles randomly switch between the two sets, it is hard to link two pseudonyms to the same vehicle, or identify a vehicle that uses solely self-certified pseudonyms within a $\Gamma$ period. Once access to the \ac{VPKI} entities is restored, $V_{5}$ refills its pseudonym pool; however, the user privacy is enhanced if it keeps switching between the two sets. In other words, if a vehicle solely relies on its \ac{VPKI}-provided pseudonyms, the probability of linking two successive pseudonyms, belonging to itself, will be increased. 

The exact threshold for how far to distribute the \acl{RHyTHM} initiation query depends on different factors, e.g., the number of nearby \ac{VPKI}-disconnected vehicles. The more vehicles without valid \ac{VPKI}-provided pseudonyms, the less is the needed support from the rest of vehicles. Clearly, initiating the \acl{RHyTHM} protocol with a high probability of $r$ to switch to self-certified pseudonyms and assist few vehicles is inefficient: it imposes extra overhead on the entire system. However, to enhance the privacy of a few users, it is sufficient to receive a small \emph{``contribution''} from other vehicles (it becomes clear later). Moreover, if the number of disconnected nodes without valid pseudonyms is much higher than the number of nodes with \ac{VPKI}-provided pseudonyms, all the nodes \emph{``should''} switch to self-certified pseudonyms, issued with aligned lifetimes. Dynamically determining an optimal $r$ remains as future work.

\vspace{-0.3em}

\section{Security \& Privacy Analysis}
\label{sec:security-analysis}

\vspace{-0.3em}

\textbf{Non-repudiation, authentication and integrity:} The \acl{RHyTHM} initiation is signed by a currently valid pseudonym, thus we achieve authentication and integrity. Digital signatures and pseudonyms ensure non-repudiation, thus, each entity can be held accountable for its actions. 

\textbf{Thwarting Sybil-based misbehavior:} An internal adversary could be equipped with two valid pseudonyms when \acl{RHyTHM} is active. We rely on the \emph{\aclu{HSM}} to ensure that all outgoing signatures are signed under one private key of a single valid (\ac{VPKI}- or self-certified) pseudonym. To mitigate generation of multiple self-certified pseudonyms, one can employ group signature schemes with such a feature~\cite{calandriello2011performance}.


\textbf{Revocation:} If a vehicle deviates from the security policies, it will be evicted from the system based on the underlying \ac{VPKI} operations. More precisely, the \ac{RA} interacts with the \ac{PCA}, the \ac{GM}, and the \ac{LTCA} to resolve, and possibly revoke, a misbehaving vehicle, thus, distributing the revocation list.

\textbf{Thwarting clogging \ac{DoS} attack:} \acl{RHyTHM} initiation flag, integrated in \acp{CAM}, is epidemically broadcasted. Upon reception of a \ac{CAM} with \acl{RHyTHM} initiation request, if vehicles can confirm a connection to the \ac{VPKI}, they simply ignore it (or choose a low value of $r$). Moreover, \acl{RHyTHM} only lasts while the \ac{VPKI} entities are out of reach, i.e., vehicles switch back to utilizing their \ac{VPKI}-provided pseudonyms at the end of $\Gamma$ period (if there is no more \acl{RHyTHM} initiation request).

\textbf{Honest-but-curious \ac{VPKI} entities:} Due to the separation of duty, no single \ac{VPKI} entity is able to fully de-anonymize a user or link pseudonyms over a long period of time. \acl{RHyTHM} improves privacy protection for vehicles with valid \ac{VPKI}-provided pseudonyms that participate in \acl{RHyTHM}: pseudonyms used for secure communication are partially linkable by the \ac{PCA} and partially by the \ac{GM} within a $\Gamma$. Communication with self-certified pseudonyms for vehicles without \ac{VPKI}-provided ones is linkable by the \ac{GM}. 


\begin{figure} [!t] 
	\vspace{-2em}
	\begin{center}
		\centering
		\subfloat[Baseline: 1\% disconnected]{ 
			\hspace{-0.75em} \includegraphics[trim=0cm 0cm 0.75cm 0.75cm, clip=true, totalheight=0.14\textheight,angle=0,keepaspectratio]{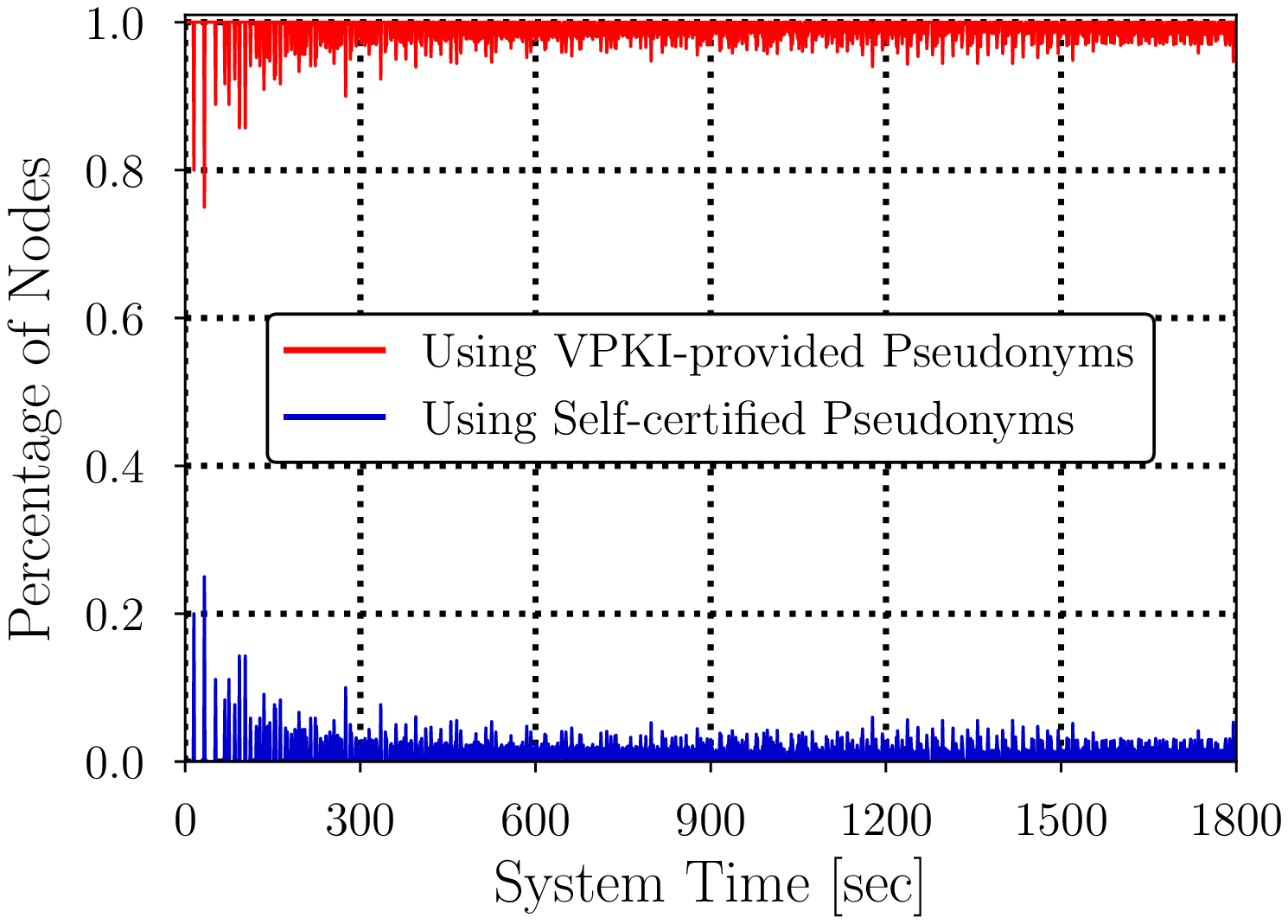}}
		\subfloat[\acl{RHyTHM}: 1\% disconnected] { 
			\hspace{-0.75em} \includegraphics[trim=0cm 0cm 0.75cm 0.75cm, clip=true, totalheight=0.14\textheight,angle=0,keepaspectratio]{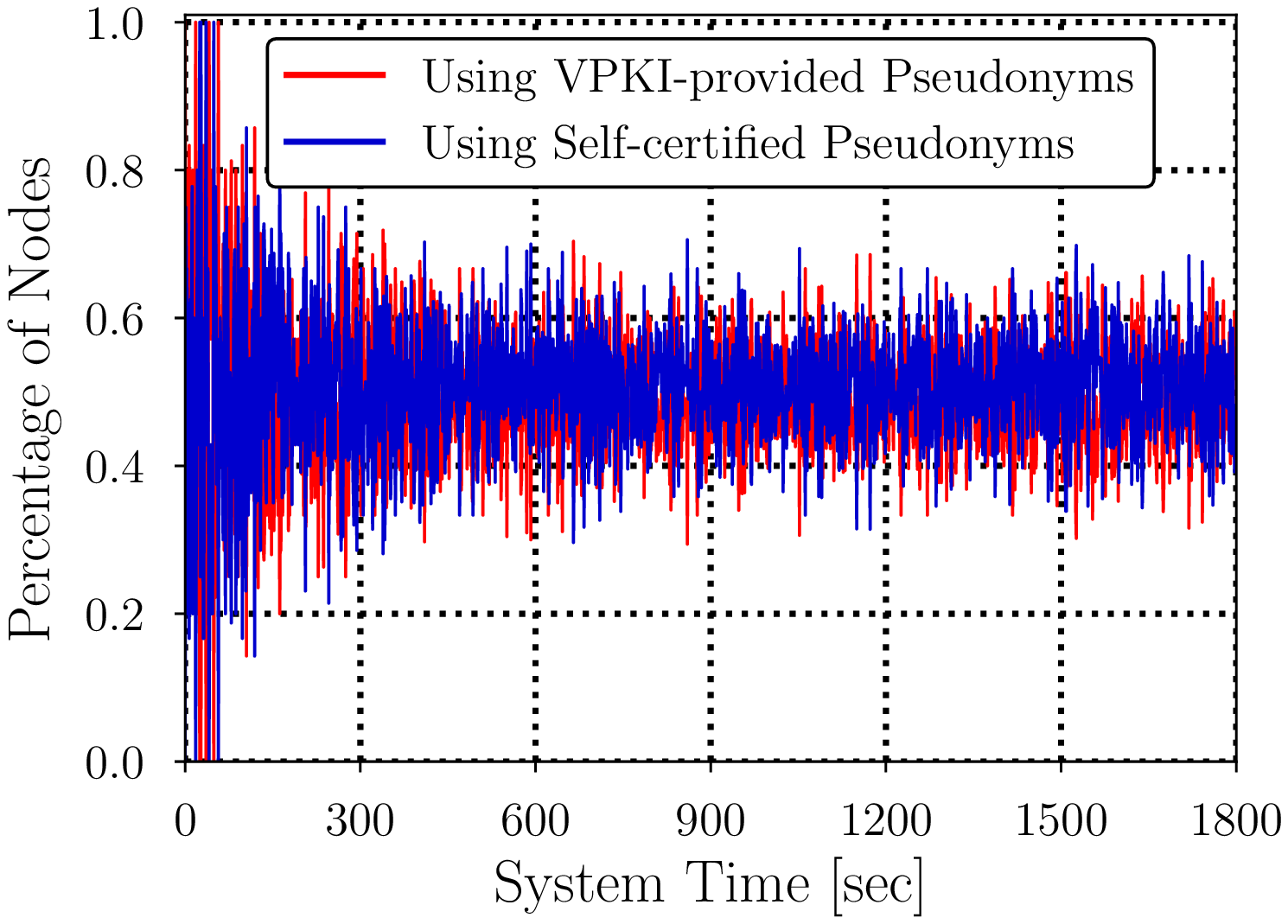}}
		\vspace{-0.35em}
		\caption{1\% of nodes run out of pseudonyms ($\tau_{P}=60$ sec, $r=0.5$)} 
		\label{fig:1-percentage-of-nodes-run-out-of-pseudonyms}
	\end{center}
	\vspace{-2.5em}
\end{figure}

\textbf{Privacy:} \acl{RHyTHM} increases user privacy in compared to the baseline scheme. We consider here a suitable privacy metric: the probability of linking two (successive) pseudonyms belonging to the same vehicle. After each pseudonym changing process, an observer might be tempted to link two pseudonyms within a region at a specific time window. Note that the \acl{RHyTHM} initiation query is signed either by a \ac{VPKI}-provided or a self-certified pseudonym(s), belonging to the disconnected node(s); thus, one can simply link that pseudonym(s) to the self-certified ones. Fig.~\ref{fig:1-percentage-of-nodes-run-out-of-pseudonyms} shows the percentage of nodes using their \ac{VPKI}-provided or self-certified pseudonyms for an actual mobility trace (www.vehicularlab.uni.lu/), during the rush hour (7-7:30 am). Fig.~\ref{fig:1-percentage-of-nodes-run-out-of-pseudonyms}.a illustrates that 1\% of nodes cannot access the \ac{VPKI} to refill their pseudonyms pool, e.g., due to sparse deployment of the \acp{RSU}. As a result, there is a huge difference between their anonymity set size, thus harming user privacy. We define the anonymity set as the set of vehicles using indistinguishable pseudonyms at any given point in time. Fig.~\ref{fig:1-percentage-of-nodes-run-out-of-pseudonyms}.b shows how \acl{RHyTHM} could enhance user privacy: nodes with valid \ac{VPKI}-provided pseudonyms randomly and independently switch to utilizing their self-certified pseudonyms to help other vehicles protect their privacy. Thus, the anonymity set size of the two groups is balanced. This does not harm the privacy of users from the larger set since they change their set randomly for each pseudonym update (it becomes clear next).

Assuming there are $N$ vehicles equipped with \ac{VPKI}-provided pseudonyms and $M$ vehicles run out of pseudonyms, thus using their self-certified pseudonyms. The probability of switching to self-certified pseudonyms is $r$. Using the baseline scheme, the probability of linking two \ac{VPKI}-provided pseudonyms belonging to the same vehicle is $\frac{1}{N}$. However, by using \acl{RHyTHM}, the probability of linking two \ac{VPKI}-provided pseudonyms belonging to the same vehicle becomes {\footnotesize  $\frac{(1-r)}{N - (r \times N)} = \frac{1}{N}$}. If a vehicle with a \ac{VPKI}-provided pseudonym decides to utilize its self-certified pseudonym in the next pseudonym update, the probability of linking those two pseudonyms becomes {\footnotesize $\frac{r}{M + (r \times N)} = \frac{1}{N + \frac{M}{r}}$}. Since {\footnotesize $\frac{1}{N + \frac{M}{r}} < \frac{1}{N}$}, if {\footnotesize $M > 0$}, the probability of linking decreases, thus enhancing user privacy. Thereby, employing \acl{RHyTHM} does not compromise the privacy of users. If a vehicle decides to utilize its self-certified pseudonym, the probability of linking decreases at the cost of extra computation overhead. Simply put, by switching back and forth between utilizing \ac{VPKI}-provided and self-certified pseudonyms, the probability of linking two pseudonyms, belonging to the same vehicle, decreases exactly because an adversary cannot know which anonymity set they belong to. If a fraction of vehicles join \acl{RHyTHM}, the probability of linking pseudonyms for those who always utilizes their \ac{VPKI}-provided pseudonyms increases exactly because an adversary should link a \ac{VPKI}-provided pseudonym to a \ac{VPKI}-provided one since the probability is higher (i.e., $\frac{1}{N} > \frac{1}{N + \frac{M}{r}}$).

\begin{figure} [!t] 
	\vspace{-3.5em}
	\begin{center}
		\centering
		\subfloat[]{
			\hspace{-1.2em} \includegraphics[trim=0.1cm 0.2cm 0.5cm 0.75cm, clip=true, 	totalheight=0.146\textheight,angle=0,keepaspectratio]{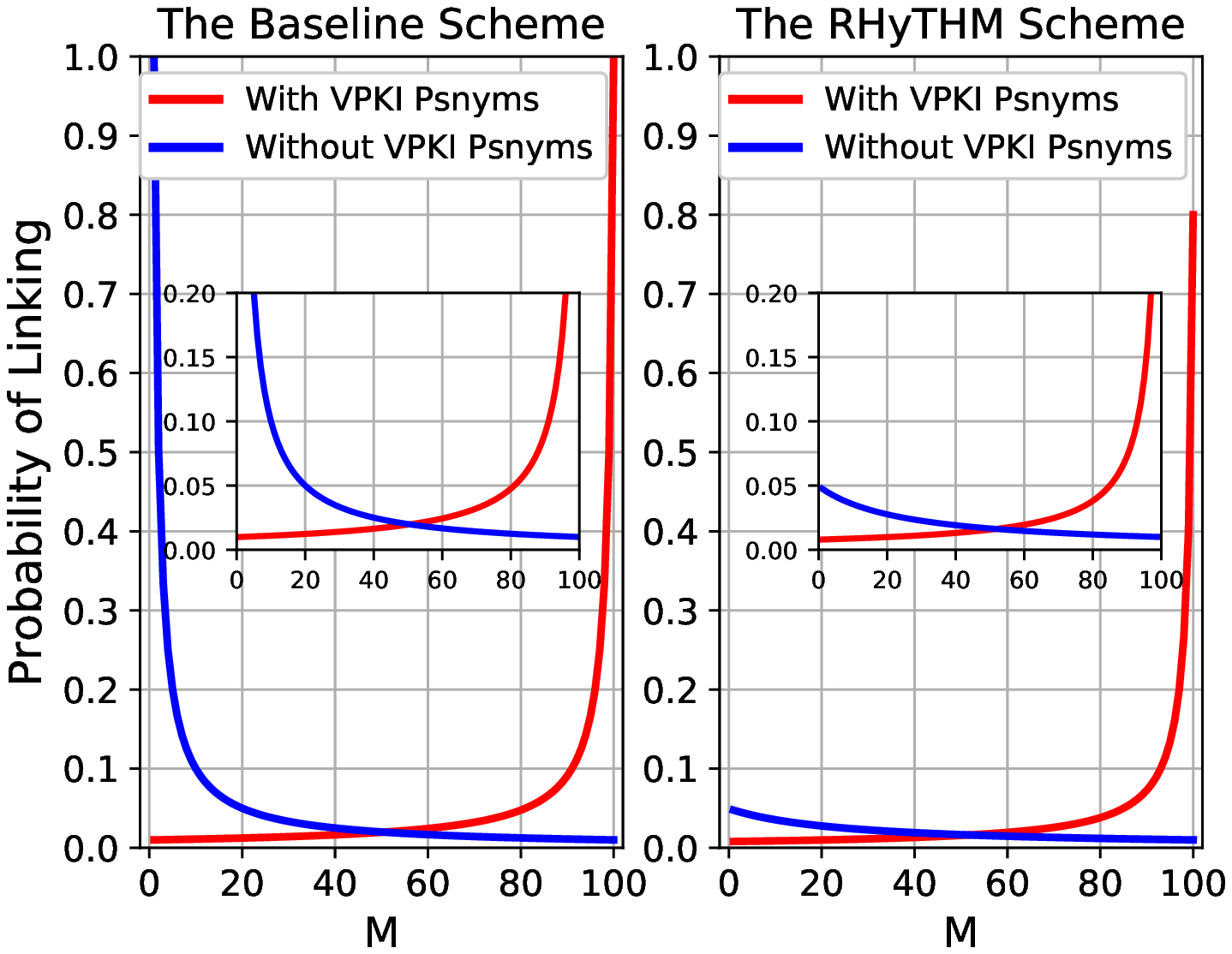}}
		\subfloat[] {
			\hspace{-1em} \includegraphics[trim=0.1cm 0.2cm 0.5cm 0.75cm, clip=true, 
			totalheight=0.146\textheight,angle=0,keepaspectratio]{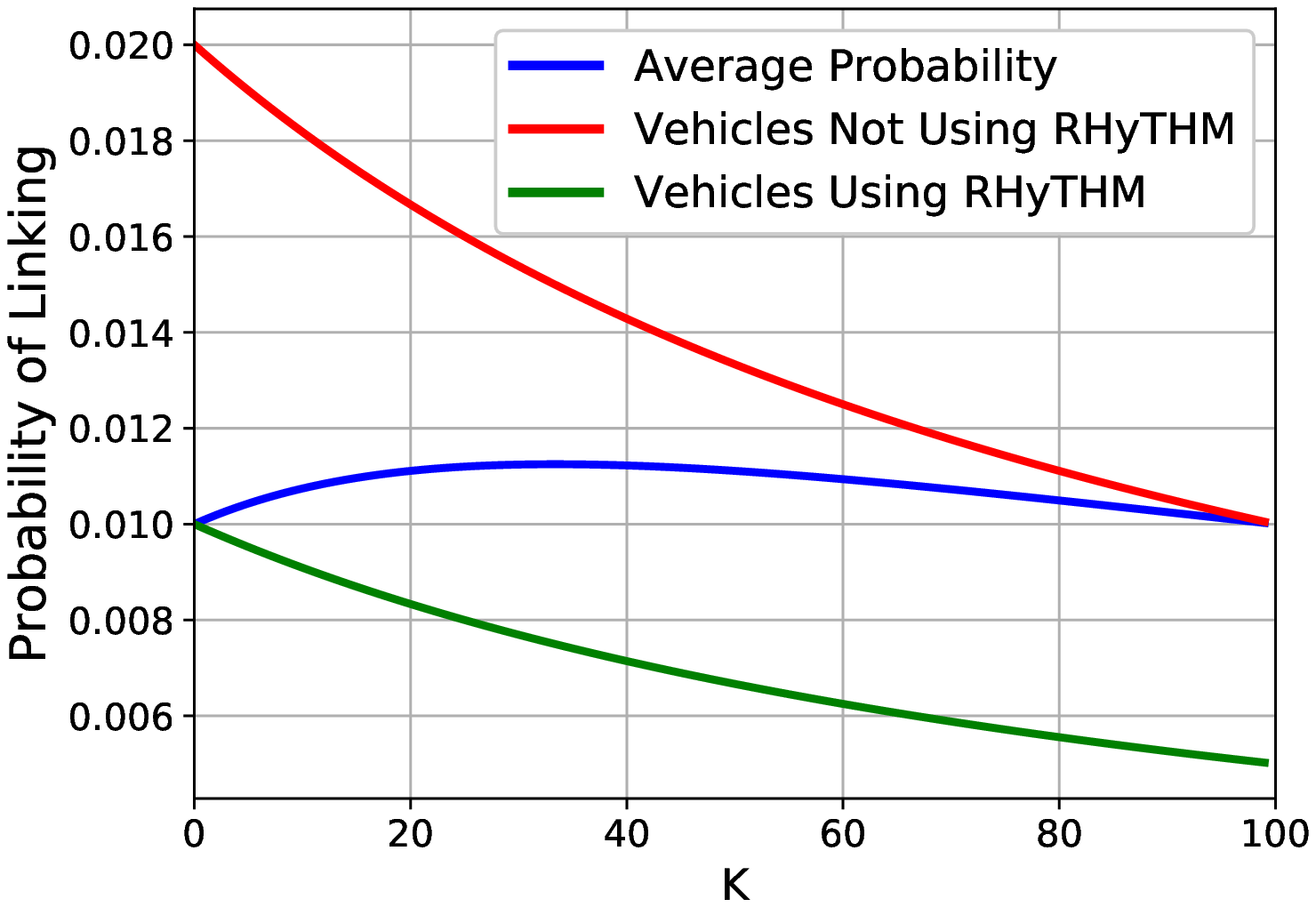}}
		\vspace{-0.37em}
		\caption{(a) Comparing the probability of linking two pseudonyms using baseline and \acl{RHyTHM} schemes ($N=100, r=0.2$). (b) Probability of linking two \ac{VPKI}-provided pseudonyms ($N=100, r=0.5$).}
		\label{fig:prob-linking-baseline-rhythm}
	\end{center}
	\vspace{-2.2em}
\end{figure}

Fig~\ref{fig:prob-linking-baseline-rhythm}.a compares the probability of linking two pseudonyms using the baseline and the \acl{RHyTHM} schemes: by employing \acl{RHyTHM}, the probability of linking self-certified pseudonyms of vehicles that must use it significantly decreases, becomes 0.05, when $M=1$ and $r=0.2$, i.e., 20 vehicles switch to their self-certified pseudonyms. Moreover, the probability of linking pseudonyms of vehicles that opt in to participate in \acl{RHyTHM} decreases slightly. When the majority of vehicles run out of pseudonyms, vehicles are highly encouraged to switch to use their self-certified pseudonyms in order to enhance their privacy. Vehicles with \ac{VPKI}-provided pseudonyms could simply ignore \acl{RHyTHM} and always use their pseudonyms. We define $K$ ({\small $0\le K \le N$}) as the number of vehicles, equipped with \ac{VPKI}-provided pseudonyms, but never join the \acl{RHyTHM} protocol. Fig.~\ref{fig:prob-linking-baseline-rhythm}.b shows that the probability of linking two \ac{VPKI}-provided pseudonyms on average, becomes:

\vspace{-1.1em}
\begin{equation*} \label{eq:probability-of-linking}
\scriptsize{	
	\begin{split}
		Pr \: \textnormal{=} \: \frac{K}{[K + (N - K) \times (1-r)]^2} + \frac{N - r \times (N - K) - K}{[K + (N - K ) \times (1-r)]^2} \times (1-r)
	\end{split}
}
\end{equation*}
\vspace{-1.1em}

The first term is the probability of linking two successive pseudonyms belonging to a vehicle not using \acl{RHyTHM}. It is the probability of the pseudonym being in $K$ set ({\footnotesize $\frac{K}{[K + (N - K)\times(1-r)]}$}), multiplied by the probability of linking it to its successive pseudonym ({\footnotesize $\frac{1}{[K + (N - K)\times(1-r)]}$}). The denominator is the size of the entire \ac{VPKI}-provided pseudonym set. The second term is for the rest of the vehicles using \acl{RHyTHM}: the probability of a pseudonym belonging to a vehicle using \acl{RHyTHM} ({\footnotesize $\frac{N - (r)\times(N - K) - K}{[K + (N - K ) \times (1-r)]}$}), multiplied by the probability of linking it to its successive pseudonym ({\footnotesize $\frac{(1-r)}{[K + (N - K )\times(1-r)]}$}). 

If {\footnotesize $K=0$}, i.e., all the vehicles use \acl{RHyTHM}, or {\footnotesize $K=N$}, i.e., the baseline scheme that vehicles with valid \ac{VPKI}-provided pseudonyms always use their pseudonyms, then the probability of linking, on average, becomes: {\footnotesize $\frac{1}{N}$}. Fig.~\ref{fig:prob-linking-baseline-rhythm}.b illustrates that using \acl{RHyTHM} increases the uncertainty as one cannot simply predict the destination set after each pseudonym update. The probability of linking two successive pseudonyms for vehicles using \acl{RHyTHM} is always less than the probability of linking for vehicles always using their \ac{VPKI}-provided pseudonyms.

\vspace{-0.5em}

\section{Performance Evaluation}
\label{sec:evaluation}

\begin{figure} [!t] 
	\vspace{-3.5em}
	\begin{center}
		\centering
		\subfloat[{\scriptsize End-to-end latency}]{
			\hspace{-1em} \includegraphics[trim=0cm 0cm 0.75cm 0.75cm, clip=true, totalheight=0.145\textheight,angle=0,keepaspectratio]{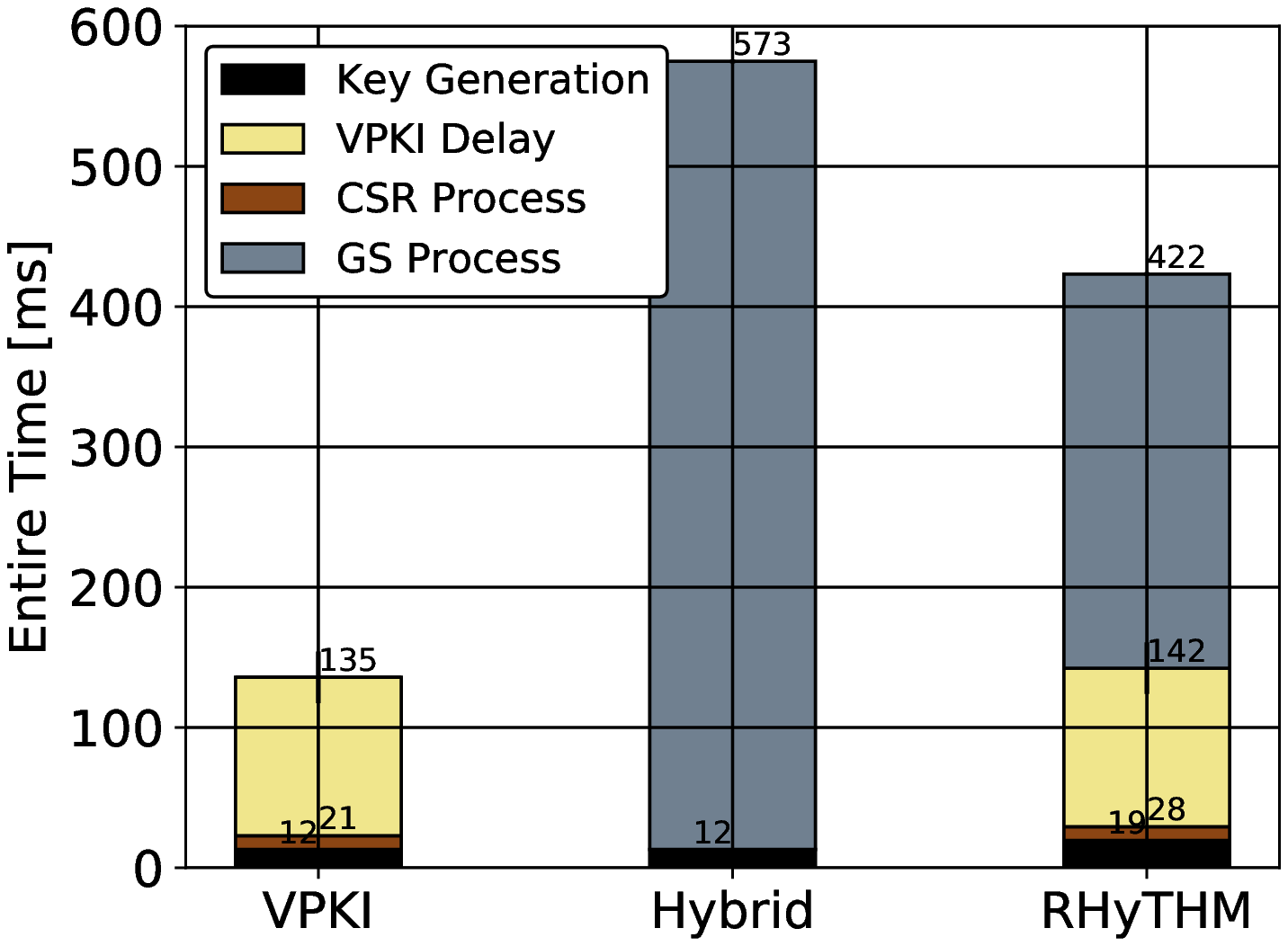}
		}
		\subfloat[{\scriptsize Cryptographic overhead}] { 
			\hspace{-1em} \includegraphics[trim=0cm 0cm 0.75cm 0.75cm, clip=true, totalheight=0.145\textheight,angle=0,keepaspectratio]{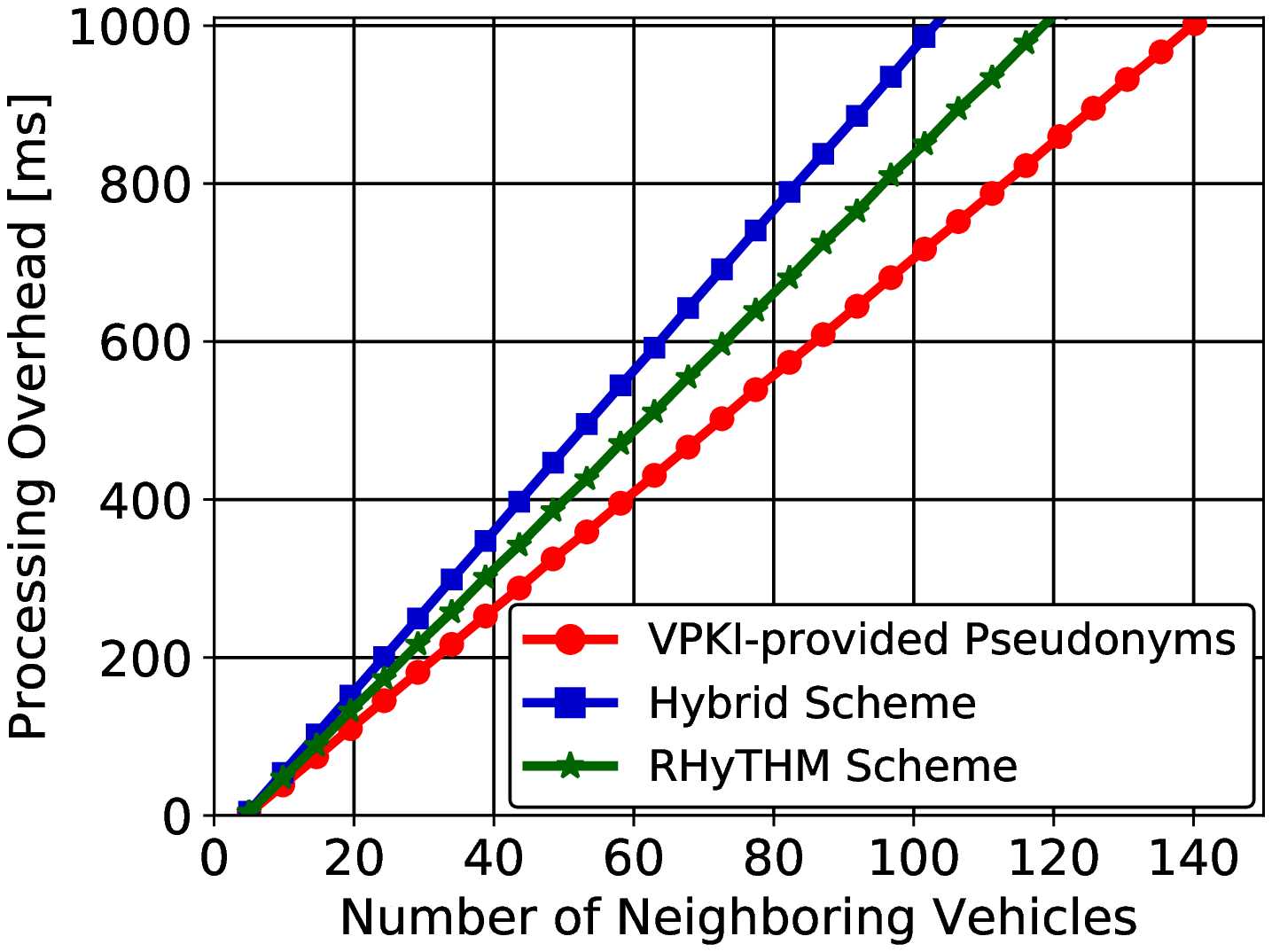}}
		\vspace{-0.25em}
		\caption{(a) End-to-end latency to acquire 10 pseudonyms, averaged over 500 runs. (b) Processing overhead as a function of the neighborhood size ($\tau_{P}=30$~sec, ratio of received messages: up to 60 beacon/sec, $r=0.5$)}
		\label{fig:communication-computation-delay}
	\end{center}
	\vspace{-2.1em}
\end{figure}

We emulate a large neighborhood with 7 Nexcom boxes (Dual-core 1.66 GHz, 1GB memory) from PRESERVE project (www.preserve-project.eu) to evaluate the performance of our scheme. Our implementation is in C, and we use OpenSSL and an implementation (github:IAIK/pairings\_in\_c) of short-group signature~\cite{boneh2004short} with security level of 112 bits for cryptographic operations and primitives. The average signing and verification latency for group signature is 56 ms and 82.5 ms, respectively; thus, the extra computation overhead, when $r=0.2$, for every vehicle in the system is around 1.6 sec per $\tau_{P}$. 

Fig.~\ref{fig:communication-computation-delay}.a shows the end-to-end latencies for obtaining 10 pseudonyms using the baseline and the \acl{RHyTHM} schemes. As the figure shows, employing \acl{RHyTHM} results in 287 ms extra overhead, mainly for generating the public/private key pairs and signing them with the $gsk$, for the vehicles equipped with valid \ac{VPKI}-provided pseudonyms. This overhead pays off as their privacy is improved compared to only using \ac{VPKI}-provided pseudonyms (beyond assisting vehicles in need). As illustrated in Fig.~\ref{fig:communication-computation-delay}.b, the total number of neighboring vehicles that an \ac{OBU} could face, if all the vehicles utilize their self-certified and \ac{VPKI}-provided pseudonyms, is 100 and 140, respectively. By employing \acl{RHyTHM} with $r=0.5$, an \ac{OBU} could verify the signature of \acp{CAM} from one up to 120 neighbors.

\vspace{-0.5em}

\section{Conclusion and Future Work}
\label{sec:conclusions}

We presented \acl{RHyTHM} as a privacy-preserving scheme to help vehicles operate and protect their privacy even if they run out of pseudonyms. As future work, we plan to investigate the provision of incentives for the vehicles to participants in \acl{RHyTHM} and the optimal probability of switching to utilizing self-certified pseudonyms in different circumstances.

\bibliographystyle{IEEEtran}
\bibliography{IEEEabrv,references}

\end{document}